\documentclass[12pt]{article}

\author{Yu.~M.~Zinoviev
       \thanks{E-mail address: Yurii.Zinoviev@ihep.ru} \\
        {\it Institute for High Energy Physics} \\
        {\it Protvino, Moscow Region, 142280, Russia}}
\title{Frame-like gauge invariant formulation \\
for massive high spin particles}

\date{}

\begin{document}

\maketitle

\begin{abstract}
In this paper we extend a so called frame-like formulation of massless
high spin particles to massive case. We start with two explicit
examples of massive spin 2 and spin 3 particles and then construct
gauge invariant description for arbitrary integer spin case.
Similarly, for the fermionic case we start with first non-trivial
example --- massive spin 5/2 particle and then construct gauge
invariant description for arbitrary half-integer spin case. In all
cases we consider massive particles in $(A)dS$ spaces with arbitrary
cosmological constant (including flat Minkowski space) and this allows
one to investigate all possible massless and partially massless limits
for such particles. 
\end{abstract}

\thispagestyle{empty}
\newpage
\setcounter{page}{1}

\section*{Introduction}

As is well known, basically there are two approaches for description
of gravity theory --- metric one, where the main object is symmetric
metric tensor $g_{\mu\nu}$, and tetrad one with tetrad $e_\mu{}^a$ and
Lorentz connection $\omega_\mu{}^{ab}$. To a large extent these two
approaches are equivalent, but for some concrete task one or another
approach can more convenient. In particular, to describe interactions
of fermionic fields with gravity (e.g. supergravity) one is forced to
use tetrad formulation. Physically, the existence of two such
approaches means that for a description of massless spin 2 particles
one can use Lagrangians of second or first order in derivatives.

These two approaches admit natural generalization for description of
high spin particles (for recent review of high spin theories see e.g.
\cite{Vas04,Sor04,BCIV05,FT08}). Generalization of metric approach has
been constructed in \cite{SH74,SH74a,Fro78,FF78,Fro79,FF80},
while generalization of tetrad approach, the so called frame-like
formalism, has been constructed in \cite{Vas80,LV88,Vas88}
(see also \cite{AD80,Zin03,Zin03a,ASV03,ASV05,SV06,ASV06,SV08,Skv08}).

As is well known, Lorentz covariant description of massless high spin
fields requires a theory to be gauge invariant. This, in particular,
lead to so called constructive approach to investigation of consistent
interactions of such fields when interaction Lagrangians and
appropriate gauge transformations are constructed iteratively by the
number of fields. In turn, common description of massive fields
requires that some constraints must follow from equation of motion
excluding all unphysical degrees of freedom. In this, at least two
general problems appear then one tries to switch on interactions.
First of all, a number of constraints could change thus leading to a
change in a number of degrees of freedom and reappearing of unphysical
ones. At second, even if a number of constraints remains to be the
same as in free theory, interacting theory very often turns out to be
non-casual, i.e. has solution corresponding to non-luminal
propagation.
 
One of the possible solutions is to use gauge invariant description of
massive high spin fields. There at least two basic approaches to such
description. One of them based on the powerful BRST method
\cite{BK05,BKRT06,BKL06,BKR07,MR07,BKT07}. Another one appeared in
attempt to generalize to high spins a very well known mechanism of
spontaneous gauge symmetry breaking \cite{Zin83,KZ97,Zin01,Met06} (see
also \cite{Zin02a,Zin03a,Med03,BHR05,HW05}). In such a breaking a set
of Goldstone fields with non-homogeneous gauge transformations appear
making gauge invariant description of massive gauge fields possible. 
Such gauge invariant description of massive fields works well not only
in flat Minkowski space-time, but in (anti) de Sitter space-times as
well. All that one needs to do is to replace ordinary partial
derivatives with the covariant ones and take into account commutator
of these derivatives which is non-zero now. In particular, this
formulation turns out to be very convenient for investigation of so
called partially massless theories which appear in de Sitter space
\cite{DW01,DW01a,DW01c,Zin01,SV06}.

It is evident that in any theory of high spin particles most of them
have to be massive (and their gauge symmetries have to be
spontaneously broken). It means that in any supersymmetric high spin
theory like the superstring these particles must belong to some
massive supermultiplet. It may seems strange but though explicit
realization of massless supermultiplets with arbitrary spins were
known for a long time \cite{Cur79} explicit construction for massive
supermultiplets was not available until recently \cite{Zin07a} (see
also \cite{Zin02,Zin07}). The main idea is that massive supermultiplet
must be easily constructed out of the appropriate set of massless ones
exactly in the same way as massive particle could be constructed using
appropriate set of massless ones.

Construction of consistent high spin particles interactions is one of
the old, hard and still unsolved problems. For the massless particles
it is possible to formulate constructive approach to this problem (for
BRST formulation see \cite{BFPT06}). In this approach one starts with
free Lagrangian for the collection of massless fields with appropriate
gauge transformations and tries to construct interacting Lagrangian
and modified gauge transformations iteratively by the number of fields
so that:
$$
{\cal L} \sim {\cal L}_0 + {\cal L}_1 + {\cal L}_2 + \dots, \qquad
\delta \sim \delta_0 + \delta_1 + \delta_2 + \dots
$$
where ${\cal L}_1$ --- cubic vertex, ${\cal L}_2$ --- quartic one and
so on, while $\delta_1$ --- corrections to gauge transformations
linear in fields, $\delta_2$ --- quadratic in fields and so on.
The mere existence of gauge invariant formulation for massive high
spin particles allows us to extend such constructive approach for any
collection of massive and/or massless particles, see e.g.
\cite{Zin06,Zin08}.
 
Till now, in most of the works on gauge invariant description for
massive high spin particles metric-like formulation was used (see,
however, \cite{Zin03a,SV06}). The aim of this paper is to extend
frame-like formulation of bosonic and fermionic high spin particles to
massive case, in this we will follow a minimalistic approach
\cite{KZ97,Zin01,Met06} introducing only minimal number of fields
which are absolutely necessary for gauge invariant description of
massive particle. We start in section 1 and section 2 with two
explicit examples of massive spin 2 and spin 3 particles
correspondingly. Then in section 3 we construct gauge invariant
description of massive particles with arbitrary integer spin, which in
metric-like formalism corresponds to completely symmetric tensors.
Similarly, in section 4 we consider explicit construction of first
non-trivial fermionic case --- massive spin 5/2 particle, and then in
section 5 we construct its generalization to arbitrary half-integer
spin particles, which in metric-like formalism corresponds to
completely symmetric spin-tensors.

\section{Spin 2}

Frame-like formalism is a very natural generalization of well known
tetrad formulation of gravity on the case of high spin fields.
We have already considered massive spin 2 particles in such
formalism \cite{Zin03a}, however, it is instructive to look how this
construction works in this simplest case before turning to
generalizations on higher spins.

Two main objects in this case are physical field $h_\mu{}^a$ and
auxiliary field $\omega_\mu{}^{ab}$, antisymmetric on $ab$. Massless
theory has to be invariant under two gauge transformations:
\begin{equation}
\delta h_\mu{}^a = \partial_\mu \xi^a + \eta_\mu{}^a, \qquad
\delta \omega_\mu{}^{ab} = \partial_\mu \eta^{ab}
\end{equation}
In this, it is easy to construct a tensor ("torsion") $T_{\mu\nu}{}^a$
which is invariant under $\xi^a$-transforma\-tions:
\begin{equation}
T_{\mu\nu}{}^a = \partial_\mu h_\nu{}^a - \partial_\nu h_\mu{}^a =
\partial_{[\mu} h_{\nu]}{}^a
\end{equation}
To construct gauge invariant Lagrangian for massless field one can
use the following simple trick. Let us consider an expression:
$$
\left\{ \phantom{|}^{\mu\nu\alpha}_{abc} \right\}
\omega_\mu{}^{ab} T_{\nu\alpha}{}^c, \qquad
\left\{ \phantom{|}^{\mu\nu\alpha}_{abc} \right\} =
\delta_a{}^{[\mu} \delta_b{}^\nu \delta_c{}^{\alpha]}
$$
and make a substitution $T_{\mu\nu}{}^a \rightarrow 
\omega_{[\mu,\nu]}{}^a$. This gives an expression
$ \left\{ \phantom{|}^{\mu\nu}_{ab} \right\}
\omega_\mu{}^{ac} \omega_\nu{}^{bc} $.
Now we look for massless Lagrangian in the form
$$
{\cal L}_0 = a_1
\left\{ \phantom{|}^{\mu\nu}_{ab} \right\}
\omega_\mu{}^{ac} \omega_\nu{}^{bc} + a_2
\left\{ \phantom{|}^{\mu\nu\alpha}_{abc} \right\}
\omega_\mu{}^{ab} T_{\nu\alpha}{}^c
$$
This Lagrangian is invariant (by construction) under
$\xi^a$-transformations, while invariance under
$\eta^{ab}$-transformations requires $a_1 = - 2 a_2$. In what follows
we choose $a_1 = \frac{1}{2}$, so finally we get:
\begin{equation}
{\cal L}_0 = \frac{1}{2}
\left\{ \phantom{|}^{\mu\nu}_{ab} \right\}
\omega_\mu{}^{ac} \omega_\nu{}^{bc} - \frac{1}{4}
\left\{ \phantom{|}^{\mu\nu\alpha}_{abc} \right\}
\omega_\mu{}^{ab} T_{\nu\alpha}{}^c
\end{equation}

Now let us switch to constant curvature $(A)dS_d$ space. Here and in
what follows we will use the following convention on $(A)dS_d$
covariant derivatives:
\begin{equation}
[ D_\mu, D_\nu ] \xi^a = - \kappa (e_\mu{}^a \xi_\nu - e_\nu{}^a
\xi_\mu ), \qquad \kappa = \frac{2 \Lambda}{(d-1)(d-2)}
\end{equation}
Now, if we replace all partial derivatives in the Lagrangian and gauge
transformations by the $(A)dS_d$-covariant ones, our massless
Lagrangian looses its gauge invariance:
$$
\delta {\cal L}_0 = \kappa (d-2) [- \omega^a \xi_a + \eta^{ab} h_{ab}]
$$
where $\omega^a = \omega_\mu{}^{\mu a}$, but invariance could be
restored by adding appropriate corrections to the Lagrangian and gauge
transformations:
\begin{equation}
\Delta {\cal L}_0 = - \frac{\kappa (d-2)}{2}
\left\{ \phantom{|}^{\mu\nu}_{ab} \right\}
h_\mu{}^a h_\nu{}^b, \qquad
\delta \omega_\mu{}^{ab} = \kappa e_\mu{}^{[a} \xi^{b]}
\end{equation}

Now let us consider a massive spin 2 particle in $(A)dS_d$ space. As
is well known and we have already seen above, even for massless
particles in $(A)dS_d$ space gauge invariance requires introduction of
mass-like terms into Lagrangians as well as appropriate corrections
for gauge transformations. So working with massive particles in
$(A)dS_d$ spaces it turns out to be very convenient to organize a
calculations by dimensionality of variations. Let us in general denote
physical field as $\Phi$, auxiliary field as $\Omega$, parameters of
gauge transformations as $\xi$ and that of local shifts as $\eta$.
Then general structure of massive Lagrangian is ${\cal L} = {\cal L}_0
+ {\cal L}_1 + {\cal L}_2$, where ${\cal L}_0 \sim \Omega \Omega
\oplus \Omega D \Phi$, ${\cal L}_1 \sim m \Omega \Phi$ and ${\cal L}_2
\sim m^2 \Phi \Phi$. At the same time, general form of gauge
transformations looks like $\delta = \delta_0 + \delta_1 + \delta_2$
where $\delta_0 \Phi \sim D \xi + \eta$, $\delta_0 \Omega \sim D
\eta$, $\delta_1 \Phi \sim m \xi$, $\delta_1 \Omega \sim m \eta$ and
$\delta_2 \Omega \sim m^2 \xi$.

As is known \cite{KZ97,Zin01}, to construct gauge invariant
formulation for massive spin 2 field we need two additional physical
fields with spin 1 and spin 0. It is natural to use first order
formalism for these fields also, so we introduce two pairs ($F^{ab},
A_\mu$) and ($\pi^a, \varphi$) and start with the sum of
(covariantized) kinetic terms for all three pairs:
\begin{equation}
{\cal L}_0 = \frac{1}{2} \left\{ \phantom{|}^{\mu\nu}_{ab} \right\}
\omega_\mu{}^{ac} \omega_\nu{}^{bc} - \frac{1}{4} \left\{
\phantom{|}^{\mu\nu\alpha}_{abc} \right\} \omega_\mu{}^{ab}
T_{\nu\alpha}{}^c + \frac{1}{4} F_{ab}{}^2 - \frac{1}{2} \left\{
\phantom{|}^{\mu\nu}_{ab} \right\} F^{ab} D_\mu A_\nu - \frac{1}{2}
\pi_a{}^2 + \left\{ \phantom{|}^{\mu}_{a} \right\} \pi^a D_\mu \varphi
\end{equation}
and initial gauge transformations:
\begin{equation}
\delta_0 h_\mu{}^a = D_\mu \xi^a, \qquad
\delta_0 \omega_\mu{}^{ab} = D_\mu \eta^{ab}, \qquad
\delta_0 A_\mu = D_\mu \xi
\end{equation}
Due to non-commutativity of $(A)dS$ covariant derivatives  this
Lagrangian is not invariant under initial gauge transformations:
$$
\delta {\cal L}_0 = \kappa (d-2) [- \omega^a \xi_a + \eta^{ab} h_{ab}]
$$
so we have to take this non-invariance into account at appropriate
stage of calculations. Now we add additional terms of order $m$ to the
Lagrangian:
\begin{equation}
{\cal L}_1 = a_1 \left\{ \phantom{|}^{\mu\nu}_{ab} \right\}
\omega_\mu{}^{ab} A_\nu + a_2 \left\{ \phantom{|}^\mu_a \right\}
F^{ab} h_\mu{}^b + a_3 \left\{ \phantom{|}^\mu_a \right\}
\pi^a A_\mu 
\end{equation}
Non-invariance of these terms under initial gauge transformations
$\delta_0 {\cal L}_1$ could be compensated by
\begin{equation}
\delta_1 h_\mu{}^a = \frac{2\alpha_1}{d-2} e_\mu{}^a \xi, \qquad
\delta_1 F^{ab} = - 2 \alpha_1 \eta^{ab}, \qquad
\delta_1 A_\mu = \alpha_1 \xi_\mu, \qquad
\delta_1 \varphi = \alpha_2 \xi
\end{equation}
provided $a_1 = a_2 = \alpha_1$, $a_3 = - \alpha_2$. Thus we obtain
$\delta_0 {\cal L}_1 + \delta_1 {\cal L}_0 = 0$ and this leaves us
with variations of order $m^2$ (taking into account non-invariance
due to non-commutativity of covariant derivatives) 
$ \delta_0 {\cal L}_0 + \delta_1 {\cal L}_1 $.
To compensate, we introduce mass-like terms into the Lagrangian as
well as appropriate corrections to gauge transformations:
\begin{equation}
{\cal L}_2 = b_1 \left\{ \phantom{|}^{\mu\nu}_{ab} \right\}
h_\mu{}^a h_\nu{}^b + b_2 \left\{ \phantom{|}^\mu_a \right\}
h_\mu{}^a \varphi + b_3 \varphi^2 
\end{equation}
\begin{equation}
\delta_2 \omega_\mu{}^{ab} = \beta_1 e_\mu{}^{[a} \xi^{b]} \qquad
\delta_2 \pi^a = \beta_2 \xi^a 
\end{equation}
The requirement that total Lagrangian be invariant under the total
gauge transformations allows one to express all parameters in terms of
$\alpha_1$ and $\alpha_2$:
$$
b_1 = \alpha_1{}^2 - \frac{\kappa (d-2)}{2}, \qquad 
\beta_1 = - \frac{2 b_1}{d-2}, \qquad 
b_2 = \beta_2 = - \alpha_1 \alpha_2, \qquad
 b_3 = \frac{d}{d-2} \alpha_1{}^2
$$
and gives important relation on these two parameters:
$$
4 (d-1) \alpha_1{}^2 - (d-2) \alpha_2{}^2 = 2 \kappa (d-1) (d-2)
$$

There is no strict definition on what is mass in $(A)dS$ spaces (see
for example discussion in \cite{Gar03}), but working with gauge
invariant formulation of massive particles it is natural to define
massless limit as a limit where all Goldstone fields decouple from the
main gauge field. For the case at hands, it means that it is the limit
$\alpha_1 \rightarrow 0$ corresponds to massless one. As for the
concrete normalization, we will follow the rule "mass is the parameter
that would be mass in flat Minkowski space". So we put $ \alpha_1{}^2
= m^2/2$. Combining all pieces together, we obtain final Lagrangian:
\begin{eqnarray}
{\cal L} &=& \frac{1}{2} \left\{ \phantom{|}^{\mu\nu}_{ab} \right\}
\omega_\mu{}^{ac} \omega_\nu{}^{bc} - \frac{1}{4} \left\{
\phantom{|}^{\mu\nu\alpha}_{abc} \right\} \omega_\mu{}^{ab}
T_{\nu\alpha}{}^c + \frac{1}{4} F_{ab}{}^2 - \frac{1}{2} \left\{
\phantom{|}^{\mu\nu}_{ab} \right\} F^{ab} D_\mu A_\nu - \frac{1}{2}
\pi_a{}^2 + \left\{ \phantom{|}^{\mu}_{a} \right\} \pi^a D_\mu \varphi
 \nonumber \\
 && + \frac{m}{\sqrt{2}} [ \left\{ \phantom{|}^{\mu\nu}_{ab} \right\}
\omega_\mu{}^{ab} A_\nu +  \left\{ \phantom{|}^\mu_a \right\}
F^{ab} h_\mu{}^b ] - \alpha_2 \left\{ \phantom{|}^\mu_a \right\}
\pi^a A_\mu + \nonumber \\
 && + \frac{m^2 - \kappa (d-2)}{2} \left\{
\phantom{|}^{\mu\nu}_{ab}
\right\} h_\mu{}^a h_\nu{}^b - \frac{m\alpha_2}{\sqrt{2}} \left\{
\phantom{|}^\mu_a \right\} h_\mu{}^a \varphi + \frac{d}{2(d-2)} m^2
\varphi^2 
\end{eqnarray}
This Lagrangian is invariant under the following gauge
transformations:
\begin{eqnarray}
\delta h_\mu{}^a &=& D_\mu \xi^a + \frac{m\sqrt{2}}{d-2} e_\mu{}^a
\xi, \qquad \delta \omega_\mu{}^{ab} = D_\mu \eta^{ab} -
(\frac{m^2}{d-2} - \kappa) e_\mu{}^{[a} \xi^{b]} \nonumber \\
\delta A_\mu &=& D_\mu \xi + \frac{m}{\sqrt{2}} \xi_\mu, \qquad \delta
F^{ab} = - m \sqrt{2} \eta^{ab}, \qquad \delta \varphi = \alpha_2 \xi,
\qquad \delta \pi^a = - \frac{m\alpha_2}{\sqrt{2}} \xi^a
\end{eqnarray}
where parameter $\alpha_2$ is defined through the relation:
\begin{equation}
(d-2) \alpha_2{}^2 = 2 (d-1) [ m^2 - \kappa (d-2) ]
\end{equation}
From the last relation one can see that in $dS$ space ($\kappa > 0$)
there is a unitary forbidden region $m^2 < \kappa (d-2)$
\cite{Hig87,Hig87a}. One can assume that it may be some deficiency of
gauge invariant description, but it is easy to construct gauge
invariant description for this region provided one change the sign of
scalar field $\varphi$ kinetic terms so that this field becomes a
ghost. Thus a massless limit is possible in $AdS$ space only, in this
vector and scalar fields decouple and describe massive spin 1
particle. On the other hand, in the $dS$ space one can put $\alpha_2 =
0$. In this, scalar field completely decouples, while two other ones
gives gauge invariant description of so called partially massless spin
2 particles with the Lagrangian:
\begin{eqnarray}
{\cal L} &=& \frac{1}{2} \left\{ \phantom{|}^{\mu\nu}_{ab} \right\}
\omega_\mu{}^{ac} \omega_\nu{}^{bc} - \frac{1}{4} \left\{
\phantom{|}^{\mu\nu\alpha}_{abc} \right\} \omega_\mu{}^{ab}
T_{\nu\alpha}{}^c + \frac{1}{4} F_{ab}{}^2 - \frac{1}{2} \left\{
\phantom{|}^{\mu\nu}_{ab} \right\} F^{ab} D_\mu A_\nu +
 \nonumber \\
 && + \frac{m}{\sqrt{2}} [ \left\{ \phantom{|}^{\mu\nu}_{ab} \right\}
\omega_\mu{}^{ab} A_\nu +  \left\{ \phantom{|}^\mu_a \right\}
F^{ab} h_\mu{}^b ] 
\end{eqnarray}
(note the absence of explicit mass terms) which is invariant under the
following gauge transformations:
\begin{eqnarray}
\delta h_\mu{}^a = D_\mu \xi^a + \frac{m\sqrt{2}}{d-2} e_\mu{}^a
\xi, && \delta \omega_\mu{}^{ab} = D_\mu \eta^{ab} -
(\frac{m^2}{d-2} - \kappa) e_\mu{}^{[a} \xi^{b]} \nonumber \\
\delta A_\mu = D_\mu \xi + \frac{m}{\sqrt{2}} \xi_\mu, && \delta
F^{ab} = - m \sqrt{2} \eta^{ab}
\end{eqnarray}

\section{Spin 3 }

In this section we consider less trivial spin 3 case which will show
some important features of arbitrary spin case. For the description of
massless spin 3 particles one needs \cite{Vas80,LV88} main field
$\Phi_\mu{}^{ab}$, which is symmetric and traceless on "local" indices
$ab$, and auxiliary field $\Omega_\mu{}^{a,bc}$ symmetric on $bc$,
traceless on all local indices and satisfying the relation
$\Omega_\mu{}^{(a,bc)} = 0$, where round brackets denote
symmetrization. To describe right number of physical degrees of
freedom, the theory must be invariant under the following gauge
transformations:
\begin{equation}
\delta \Phi_\mu{}^{ab} = \partial_\mu \xi^{ab} + \eta_\mu{}^{ab},
\qquad \delta \Omega_\mu{}^{a,bc} = \partial_\mu \eta^{a,bc} +
\zeta_\mu{}^{a,bc}
\end{equation}
where $\xi^{ab}$ symmetric and traceless, $\eta^{a,bc}$ has the
same properties as $\Omega$, while $\zeta^{ab,cd}$ is symmetric on
$ab$ and $cd$, traceless on all indices and satisfies a constraint
$\zeta^{(ab,c)d} = 0$. Note that in general one also needs so called
extra filed $\Omega_\mu{}^{ab,cd}$ having the same properties on local
indices as $\zeta$ and playing the role of gauge field for $\zeta$
transformations. However, such extra fields do not enter free
Lagrangians (though they play important role in construction of
interactions) so we will not introduce such fields in this work.
 As in the previous case, one can easily construct an object out of
first derivatives of $\Phi$ which will be invariant under
$\xi$-transformations $ T_{\mu\nu}{}^{ab} = \partial_{[\mu}
\Phi_{\nu]}{}^{ab} $. To find a correct structure of massless
Lagrangian one can use the same trick:
$$
\left\{ \phantom{|}^{\mu\nu\alpha}_{abc} \right\} \Omega_\mu{}^{a,bd}
T_{\nu\alpha}{}^{cd} \quad \rightarrow \quad 2 \left\{
\phantom{|}^{\mu\nu\alpha}_{abc}
\right\} \Omega_\mu{}^{a,bd} \Omega_{\nu,\alpha}{}^{cd} = 
\left\{ \phantom{|}^{\mu\nu}_{ab} \right\} [ \Omega_\mu{}^{a,cd}
\Omega_\nu{}^{b,cd} + 2 \Omega_\mu{}^{c,ad} \Omega_\nu{}^{c,bd} ]
$$
So we will look for the appropriate Lagrangian in the form:
$$
{\cal L}_0 = a_1 \left\{ \phantom{|}^{\mu\nu}_{ab} \right\} [
\Omega_\mu{}^{a,cd} \Omega_\nu{}^{b,cd} + 2 \Omega_\mu{}^{c,ad}
\Omega_\nu{}^{c,bd} ] + a_ 2 \left\{ \phantom{|}^{\mu\nu\alpha}_{abc}
\right\} \Omega_\mu{}^{a,bd} T_{\nu\alpha}{}^{cd}
$$
This Lagrangian is invariant under $\xi$-transformations, while
invariance under $\eta$-transformations requires $a_2 = - 2 a_1$
(in this, the Lagrangian is invariant under $\zeta$-transformations as
well). We choose $a_1 = - 1/6$, $a_2 = 1/3$ and obtain finally:
\begin{equation}
{\cal L}_0 = - \frac{1}{6} \left\{ \phantom{|}^{\mu\nu}_{ab} \right\}
[ \Omega_\mu{}^{a,cd} \Omega_\nu{}^{b,cd} + 2 \Omega_\mu{}^{c,ad}
\Omega_\nu{}^{c,bd} ] + \frac{1}{3} \left\{
\phantom{|}^{\mu\nu\alpha}_{abc} \right\} \Omega_\mu{}^{a,bd}
T_{\nu\alpha}{}^{cd}
\end{equation}

Now let us consider deformation to $(A)dS$ space \cite{LV88}. If one
replaces all derivatives in the Lagrangian and gauge transformations
by the $(A)dS$-covariant ones, the Lagrangian cease to be invariant:
$$
\delta {\cal L}_0 = \kappa (d-1) [ \Omega_\mu{}^{\mu,ab} \xi_{ab} -
\eta^{a,bc} \Phi_{a,bc} ]
$$
but this non-invariance could easily be compensated by adding
appropriate corrections to the Lagrangian and gauge transformations:
\begin{equation}
\Delta {\cal L}_0 = \kappa (d-1) \left\{ \phantom{|}^{\mu\nu}_{ab}
\right\} \Phi_\mu{}^{ac} \Phi_\nu{}^{bc}
\end{equation}
\begin{equation}
\delta \Omega_\mu{}^{a,bc} = \frac{d-1}{d-2} \kappa [ (2 e_\mu{}^a
\xi^{bc} - e_\mu{}^{(b} \xi^{c)a}) + \frac{1}{d-1} (
2 g^{bc} \xi_\mu{}^a - g^{a(b} \xi_\mu{}^{c)}]
\end{equation}

Let us turn to the massive case. Now we need \cite{KZ97,Zin01} three
additional physical fields, corresponding to spin 2, spin 1 and spin
0. Again we will use frame-like formalism for all fields and start
with the sum of kinetic terms for all fields where all derivatives are
$(A)dS$-covariant ones:
\begin{eqnarray}
{\cal L}_0 &=& - \frac{1}{6} \left\{ \phantom{|}^{\mu\nu}_{ab}
\right\} [ \Omega_\mu{}^{a,cd} \Omega_\nu{}^{b,cd} + 2
\Omega_\mu{}^{c,ad} \Omega_\nu{}^{c,bd} ] + \frac{1}{3} \left\{
\phantom{|}^{\mu\nu\alpha}_{abc} \right\} \Omega_\mu{}^{a,bd}
T_{\nu\alpha}{}^{cd} + \frac{1}{2} \left\{ \phantom{|}^{\mu\nu}_{ab}
\right\} \omega_\mu{}^{ac} \omega_\nu{}^{bc} - \nonumber \\
 && - \frac{1}{4} \left\{ \phantom{|}^{\mu\nu\alpha}_{abc} \right\}
\omega_\mu{}^{ab} T_{\nu\alpha}{}^c + \frac{1}{4} F_{ab}{}^2 -
\frac{1}{2} \left\{ \phantom{|}^{\mu\nu}_{ab} \right\} F^{ab} D_\mu
A_\nu - \frac{1}{2} \pi_a{}^2 + \left\{ \phantom{|}^{\mu}_{a} \right\}
\pi^a D_\mu \varphi
\end{eqnarray}
as well as with full set of initial gauge transformations:
\begin{eqnarray}
\delta_0 \Phi_\mu{}^{ab} &=& D_\mu \xi^{ab} + \eta_\mu{}^{ab},
\qquad \delta_0 \Omega_\mu{}^{a,bc} = D_\mu \eta^{a,bc} \nonumber \\
\delta_0 h_\mu{}^a &=& D_\mu \xi^a, \qquad
\delta_0 \omega_\mu{}^{ab} = D_\mu \eta^{ab}, \qquad
\delta_0 A_\mu = D_\mu \xi
\end{eqnarray}
Again, due to non-commutativity of $(A)dS$ covariant derivatives this
Lagrangian is not invariant:
$$
\delta_0 {\cal L}_0 = \kappa (d-1) [ \Omega_\mu{}^{\mu,ab} \xi_{ab} -
\eta^{a,bc} \Phi_{a,bc} ] - \kappa (d-2) [ \omega^a \xi_a - \eta^{ab}
h_{ab}]
$$
but we will take this non-invariance into account later. Now we add to
the Lagrangian all possible terms of order $m$:
\begin{equation}
{\cal L}_1 = \left\{ \phantom{|}^{\mu\nu}_{ab} \right\} [ a_1
\Omega_\mu{}^{a,bc}  h_\nu{}^c + a_2 \Phi_\mu{}^{ac} \omega_\nu{}^{bc}
+ a_3 \omega_\mu{}^{ab} A_\nu ] + \left\{ \phantom{|}^{\mu}_{a}
\right\} [ a_4 h_\mu{}^b F^{ab} + a_5 A_\mu \pi^a ]
\end{equation}
As usual, non-invariance of these terms under the initial gauge
transformations $\delta_0 {\cal L}_1$ could be compensated by
appropriate corrections to gauge transformations:
\begin{eqnarray}
\delta_1 \Omega_\mu{}^{a,bc} &=& \frac{3\alpha_1}{2d} [ \eta^{a(b}
e_\mu{}^{c)} + \frac{1}{d-1} (2 g^{bc} \eta_\mu{}^a
- g^{a(b} \eta_\mu{}^{c)} ) ] \nonumber \\
\delta_1 \Phi_\mu{}^{ab} &=& \frac{3\alpha_1}{2(d-1)} (e_\mu{}^{(a}
\xi^{b)} - \frac{2}{d} g^{ab} \xi_\mu )  \\ 
\delta_1 h_\mu{}^a &=& \alpha_1 \xi_\mu{}^a + \frac{2\alpha_2}{d-2}
e_\mu{}^a \xi, \qquad \delta_1 \omega_\mu{}^{ab} = \alpha_1
\eta^{[a,b]}{}_\mu \nonumber \\
\delta_1 A_\mu &=& \alpha_2 \xi_\mu, \qquad \delta F^{ab} = - 2
\alpha_2 \eta^{ab}, \qquad \delta_1 \varphi = \alpha_3 \xi \nonumber
\end{eqnarray}
provided $a_1 = a_2 = - \alpha_1$, $a_3 = a_4 = \alpha_2$, $a_5 = -
\alpha_3$.
We proceed by adding all possible mass-like terms to the Lagrangian:
\begin{equation}
{\cal L}_2 = \left\{ \phantom{|}^{\mu\nu}_{ab} \right\} [ b_1
\Phi_\mu{}^{ac} \Phi_\nu{}^{bc} + b_2 h_\mu{}^a h_\nu{}^b ] + 
b_3 \left\{ \phantom{|}^\mu_a \right\} h_\mu{}^a \varphi + b_4
\varphi^2 
\end{equation}
as well as appropriate corrections to gauge transformations:
\begin{eqnarray}
\delta_2 \Omega_\mu{}^{a,bc} &=& \beta_1 [ (2 e_\mu{}^a \xi^{bc} -
e_\mu{}^{(b} \xi^{c)a} ) + \frac{1}{d-1} ( 2 g^{bc}
\xi_\mu{}^a - g^{a(b} \xi_\mu{}^{c)} )] \nonumber \\
\delta_2 \omega_\mu{}^{ab} &=& \beta_2 e_\mu{}^{[a} \xi^{b]}, \qquad
\delta_2 \pi^a = \beta_3 \xi^a 
\end{eqnarray}
Then cancellation of all remaining variations determines all parameters
in the Lagrangian and gauge transformations in terms of three main
ones $\alpha_{1,2,3}$:
$$
b_1 = - \frac{3\alpha_1{}^2}{2} + \kappa (d-1), \quad
\beta_1 = \frac{b_1}{d-2}, \quad
b_2 = - \frac{3(d^2-4)}{4 d (d-1)} \alpha_1{}^2 + \alpha_2{}^2 -
\frac{\kappa (d-2)}{2},
$$
$$
\beta_2 = - \frac{2 b_2}{d-2}, \quad 
b_3 = \beta_3 = - \alpha_2 \alpha_3, \quad
b_4 = \frac{d}{d-2} \alpha_2{}^2
$$
and gives two important relations on these parameters:
$$
3 (d+1) \alpha_1{}^2 - d \alpha_2{}^2 = d (d+1) \kappa, \qquad
9 d \alpha_1{}^2 - (d-2) \alpha_3{}^2 = 6 d (d-1) \kappa
$$
From this results it follows that it is the limit $\alpha_1
\rightarrow 0$ corresponds to the massless one, in this our definition
of mass gives $\alpha_1{}^2 = m^2/3$. Collecting all pieces together
we obtain final Lagrangian:
\begin{eqnarray}
{\cal L} &=& {\cal L}_0 (\Omega_\mu{}^{a,bc},\Phi_\mu{}^{ab}) + {\cal
L}_0 (\omega_\mu{}^{ab}, h_\mu{}^a) + {\cal L}_0 (F^{ab}, A_\mu) +
{\cal L}_0 (\pi^a, \varphi) - \nonumber \\
 && - \frac{m}{\sqrt{3}}  \left\{ \phantom{|}^{\mu\nu}_{ab} \right\} [
 \Omega_\mu{}^{a,bc}  h_\nu{}^c + \Phi_\mu{}^{ac} \omega_\nu{}^{bc} ]
+ \alpha_2 [ \left\{ \phantom{|}^{\mu\nu}_{ab} \right\}
 \omega_\mu{}^{ab} A_\nu + \left\{ \phantom{|}^{\mu}_{a} \right\}
h_\mu{}^b F^{ab} ] - \alpha_3 \left\{ \phantom{|}^{\mu}_{a} \right\}
A_\mu \pi^a - \nonumber \\
 &&  - \frac{m^2  - 2\kappa (d-1)}{2} \left\{
\phantom{|}^{\mu\nu}_{ab} \right\} [ \Phi_\mu{}^{ac} \Phi_\nu{}^{bc} -
\frac{3d}{2(d-1)} h_\mu{}^a h_\nu{}^b ]  - \nonumber \\
 && - \alpha_2 \alpha_3 \left\{ \phantom{|}^\mu_a \right\}
h_\mu{}^a \varphi + \frac{d}{d-2} \alpha_2{}^2 \varphi^2 
\end{eqnarray}
where parameters $\alpha_2$ and $\alpha_3$ are defined through the
relations:
\begin{equation}
d \alpha_2{}^2 = (d+1) [ m^2 - d  \kappa ], \qquad
(d-2) \alpha_3{}^2 = 3d [ m^2 - 2 (d-1) \kappa ]
\end{equation}
From the last relations we see that in $dS$ space we again obtain
unitary forbidden region $m^2 < 2 (d-1) \kappa$, so that massless
limit is possible in $AdS$ space only. The boundary of forbidden
region corresponds to $\alpha_3 = 0$, in this scalar field decouples,
while the remaining fields provides gauge invariant description of the
first partially massless theory (in $d=4$ it is a particle with
helicities $\pm 3, \pm 2, \pm 1$) with the Lagrangian (note the
absence of explicit mass-like terms):
\begin{eqnarray}
{\cal L} &=& {\cal L}_0 (\Omega_\mu{}^{a,bc},\Phi_\mu{}^{ab}) + {\cal
L}_0 (\omega_\mu{}^{ab}, h_\mu{}^a) + {\cal L}_0 (F^{ab}, A_\mu)
 - \nonumber \\
 && - \frac{m}{\sqrt{3}}  \left\{ \phantom{|}^{\mu\nu}_{ab} \right\} [
 \Omega_\mu{}^{a,bc}  h_\nu{}^c + \Phi_\mu{}^{ac} \omega_\nu{}^{bc} ]
+ \alpha_2 [ \left\{ \phantom{|}^{\mu\nu}_{ab} \right\}
 \omega_\mu{}^{ab} A_\nu + \left\{ \phantom{|}^{\mu}_{a} \right\}
h_\mu{}^b F^{ab} ]  
\end{eqnarray}
where $d \alpha_2{}^2 = (d+1)(d-2) \kappa$.

Another example of partially massless theory which "lives" in a
forbidden region we obtain by setting $m^2 = d \kappa$, i.e. $\alpha_2
= 0$. In this case both vector field $A_\mu$ and scalar one $\varphi$
decouple, while remaining $\Phi_\mu{}^{ab}$ and $h_\mu{}^a$ provides
gauge invariant description of partially massless particle (in $d=4$
it has helicities $\pm 3, \pm 2$) with the Lagrangian:
\begin{eqnarray}
{\cal L} &=& {\cal L}_0 (\Omega_\mu{}^{a,bc},\Phi_\mu{}^{ab}) + {\cal
L}_0 (\omega_\mu{}^{ab}, h_\mu{}^a)  - \frac{m}{\sqrt{3}}  \left\{
\phantom{|}^{\mu\nu}_{ab} \right\} [  \Omega_\mu{}^{a,bc}  h_\nu{}^c +
\Phi_\mu{}^{ac} \omega_\nu{}^{bc} ]  + \nonumber \\
 &&  + \frac{d-2}{2d} m^2 \left\{
\phantom{|}^{\mu\nu}_{ab} \right\} [ \Phi_\mu{}^{ac} \Phi_\nu{}^{bc} -
\frac{3d}{2(d-1)}  h_\mu{}^a h_\nu{}^b  ]
\end{eqnarray}
which is invariant under the following gauge transformations:
\begin{eqnarray}
\delta \Phi_\mu{}^{ab} &=& D_\mu \xi^{ab} + \eta_\mu{}^{ab} +
\frac{m\sqrt{3}}{2(d-1)} (e_\mu{}^{(a} \xi^{b)}
 - \frac{2}{d} g^{ab} \xi_\mu ) \nonumber \\
\delta \Omega_\mu{}^{a,bc} &=& D_\mu \eta^{a,bc} + 
\frac{m\sqrt{3}}{2d} [ \eta^{a(b} e_\mu{}^{c)} + \frac{1}{d-1} (2
g^{bc} \eta_\mu{}^a - g^{a(b} \eta_\mu{}^{c)} ] + \nonumber \\
 && + \frac{m^2}{2d} [ (2 e_\mu{}^a \xi^{bc} -
e_\mu{}^{(b} \xi^{c)a} ) + \frac{1}{d-1} ( 2 g^{bc}
\xi_\mu{}^a - g^{a(b} \xi_\mu{}^{c)})]  \\
\delta h_\mu{}^a &=& D_\mu \xi^a + \eta_\mu{}^a + \frac{m}{\sqrt{3}}
\xi_\mu{}^a, \qquad \delta \omega_\mu{}^{ab} = D_\mu \eta^{ab} +
\frac{m}{\sqrt{3}} \eta^{[a,b]}{}_\mu + \frac{3m^2}{2(d-1)}
e_\mu{}^{[a} \xi^{b]} \nonumber
\end{eqnarray}

\section{Arbitrary integer spin}

For description of massless spin $s$ particles one needs
\cite{Vas80,LV88} main physical field $\Phi_\mu{}^{a_1 \dots a_{s-1}}$
symmetric and traceless on local indices and auxiliary field
$\Omega_\mu{}^{a,a_1 \dots a_{s-1}}$ which must be symmetric on last
$s-1$ local indices, traceless on all local indices and satisfy
condition  $\Omega_{\mu}{}^{(a,a_1 \dots a_{s-1})} = 0$ (as in the
spin 3 case we will not introduce any extra fields here). To have a
correct number of physical degrees of freedom theory must be invariant
under the following gauge transformations:
\begin{equation}
\delta \Phi_\mu{}^{a_1 \dots a_{s-1}} = \partial_\mu 
\xi^{a_1 \dots a_{s-1}} + \eta_\mu{}^{a_1 \dots a_{s-1}}, \qquad
\delta \Omega_{\mu}{}^{a,a_1 \dots a_{s-1}}  = \partial_\mu 
\eta^{a,a_1 \dots a_{s-1}} + \zeta_\mu{}^{a,a_1 \dots a_{s-1}}
\end{equation}
where parameter $\xi$ --- symmetric and traceless, parameter
$\eta$ has the same properties on local indices as $\Omega$, while
$\zeta^{bc,a_1 \dots a_{s-1}}$ is symmetric on both groups of indices,
completely traceless and satisfies a constraint 
$\zeta^{b(c,a_1 \dots a_{s-1})} = 0$. As in the previous cases, we
introduce an object 
$$
T_{\mu\nu}{}^{a_1 \dots a_{s-1}} = \partial_{[\mu} 
\Phi_{\nu]}{}^{a_1 \dots a_{s-1}}
$$
invariant under $\xi$-transformations and consider an expression:
$$
\left\{ \phantom{|}^{\mu\nu\alpha}_{abc} \right\}
\Omega_\mu{}^{a,b a_2 \dots a_{s-1}} 
T_{\nu\alpha}{}^{c a_2 \dots a_{s-1}} \quad \Rightarrow \quad
\left\{ \phantom{|}^{\mu\nu\alpha}_{abc} \right\}
\Omega_\mu{}^{a,b a_2 \dots a_{s-1}}
\Omega_{\nu,\alpha}{}^{c a_2 \dots a_{s-1}} =
$$
$$
= \left\{ \phantom{|}^{\mu\nu}_{ab} \right\} [
\Omega_\mu{}^{c,a a_2 \dots a_{s-1}} 
\Omega_\nu{}^{c, b a_2 \dots a_{s-1}} + \frac{1}{s-1}
\Omega_\mu{}^{a, a_1 \dots a_{s-1}} 
\Omega_\nu{}^{b, a_1 \dots a_{s-1}} ]
$$
Let us introduce condensed notations for tensor objects like 
$\Phi_\mu{}^{a_1 \dots a_{s-1}} = \Phi_\mu{}^{(s-1)}$ and
$\Omega_\mu{}^{a,a_1 \dots a_{s-1}} = \Omega_\mu{}^{a,(s-1)}$. In this
notations our candidate for massless Lagrangian will looks as follows:
$$
{\cal L}_0 = a_1 \left\{ \phantom{|}^{\mu\nu}_{ab} \right\} [
\Omega_\mu{}^{c,a (s-2)} \Omega_\nu{}^{c, b (s-2)} + \frac{1}{s-1}
\Omega_\mu{}^{a, (s-1)} \Omega_\nu{}^{b, (s-1)} ] + a_2
\left\{ \phantom{|}^{\mu\nu\alpha}_{abc} \right\}
\Omega_\mu{}^{a,b (s-2)} T_{\nu\alpha}{}^{c (s-2)}
$$
It is (by construction) invariant under $\xi$-transformations, while
invariance under $\eta$-transforma\-tions requires $a_2 = - a_1$. For
simplicity we will use non-canonical normalization of fields and
choose the following final form for our massless Lagrangian:
\begin{equation}
(-1)^s {\cal L}_0 = \left\{ \phantom{|}^{\mu\nu}_{ab} \right\} [
\Omega_\mu{}^{c,a (s-2)} \Omega_\nu{}^{c, b (s-2)} + \frac{1}{s-1}
\Omega_\mu{}^{a, (s-1)} \Omega_\nu{}^{b, (s-1)} ] -
\left\{ \phantom{|}^{\mu\nu\alpha}_{abc} \right\}
\Omega_\mu{}^{a,b (s-2)} T_{\nu\alpha}{}^{c (s-2)}
\end{equation}
Now we consider deformation to $(A)dS$ space \cite{LV88}. If one
replaces all derivatives in the Lagrangian and gauge transformations
by $(A)dS$ covariant ones, the Lagrangian cease to be invariant:
$$
(-1)^s \delta {\cal L}_0 = \frac{2 s (d+s-4)}{s-1} \kappa [
 \Phi_\mu{}^{(s-1)} \eta^{\mu,(s-1)} - \Omega_\mu{}^{\mu,(s-1)}
\xi^{(s-1)} ]
$$
but this non-invariance could be compensated by adding appropriate
corrections to Lagrangian and gauge transformations:
\begin{equation}
(-1)^s \Delta {\cal L}_0 = - s (d+s-4) \kappa \left\{
\phantom{|}^{\mu\nu}_{ab} \right\} \Phi_\mu{}^{a(s-2)}
\Phi_\nu{}^{b(s-2)}
\end{equation}
\begin{eqnarray}
\delta \Omega_\mu{}^{a,(s-1)} &=& \frac{d+s-4}{d-2} \kappa
 \left[ (s-1) e_\mu{}^a \xi^{(s-1)} - e_\mu{}^{(1} 
\xi^{s-2)a} - \right. \nonumber \\
 && \left. - \frac{1}{d+s-4} [ (s-2) g^{a(1} \xi_\mu{}^{s-2)} - 2
g^{(12} \xi_\mu{}^{s-3)a} ] \right]
\end{eqnarray}

Let us consider massive case now. For gauge invariant description of
massive spin $s$ particle one needs a set of fields with spins $k$, $0
\le k \le s$. General formulas given above work for $k \ge 2$ only, so
spin 1 and spin 0 must be treated separately. As before, we start with
the sum of (covariantized) kinetic terms for all fields:
\begin{equation}
{\cal L}_0 = \sum_{k=2}^s {\cal L}_0(\Phi_k) + \frac{1}{4} F_{ab}{}^2
- \frac{1}{2} \left\{ \phantom{|}^{\mu\nu}_{ab} \right\} F^{ab} D_\mu
A_\nu - \frac{1}{2} \pi_a{}^2 + \left\{ \phantom{|}^{\mu}_{a} \right\}
\pi^a D_\mu \varphi
\end{equation}
where
$$
(-1)^k {\cal L}_0(\Phi_k) = \left\{ \phantom{|}^{\mu\nu}_{ab} \right\}
[ \Omega_\mu{}^{c,a (k-2)} \Omega_\nu{}^{c, b (k-2)} + \frac{1}{k-1}
\Omega_\mu{}^{a, (k-1)} \Omega_\nu{}^{b, (k-1)} ] -
\left\{ \phantom{|}^{\mu\nu\alpha}_{abc} \right\}
\Omega_\mu{}^{a,b (k-2)} T_{\nu\alpha}{}^{c (k-2)}
$$
and corresponding set of initial gauge transformations:
\begin{equation}
\delta_0 \Phi_\mu{}^{(k)} = D_\mu \xi^{(k)} + \eta_\mu{}^{(k)}, \quad
\delta_0 \Omega_\mu{}^{a,(k)} = D_\mu \eta^{a,(k)}, \quad 
1 \le k \le s-1, \qquad \delta_0 A_\mu = D_\mu \xi
\end{equation}
Due to non-commutativity of covariant derivatives our Lagrangian is
not invariant under the initial gauge transformations:
$$
\delta_0 {\cal L}_0 = \sum_{k=1}^{s-1} (-1)^k  
\frac{2 (k+1) (d+k-3)}{k} \kappa [  \Phi_\mu{}^{(k)} \eta^{\mu,(k)} -
\Omega_\mu{}^{\mu,(k)} \xi^{(k)} ]
$$
so we have to take this non-invariance into account at appropriate
stage of calculations.

Now we proceed by adding to the Lagrangian all possible terms of order
$m$:
\begin{equation}
{\cal L}_1 = \sum_{k=2}^s (-1)^k \left\{ \phantom{|}^{\mu\nu}_{ab}
\right\} [ a_k \Omega_\mu{}^{a,b(k-2)} \Phi_\nu{}^{(k-2)} + b_k
\Phi_\mu{}^{a(k-2)} \Omega_\nu{}^{b,(k-2)} ] - b_1 (A \pi)
\end{equation}
Consider gauge transformations for field $\Phi_k$. In general case ($k
\ne s$ and $k \ne 1$) this field enters ${\cal L}_1$ as follows:
$$
(-1)^k \left\{ \phantom{|}^{\mu\nu}_{ab} \right\} [ - a_{k+1}
\Omega_\mu{}^{a,b(k-1)} \Phi_\nu{}^{(k-1)} - b_{k+1}
\Phi_\mu{}^{a(k-1)} \Omega_\nu{}^{b,(k-1)} +
$$
$$
\qquad \quad + a_k \Omega_\mu{}^{a,b(k-2)} \Phi_\nu{}^{(k-2)} + b_k
\Phi_\mu{}^{a(k-2)} \Omega_\nu{}^{b,(k-2)} ]
$$
As usual, non-invariance of these terms under the initial gauge
transformations of $\Phi_k$ field could be compensated by the
following corrections to gauge transformations:
\begin{eqnarray}
\delta_1 \Phi_\mu{}^{(k)} &=&  \frac{k \alpha_k}{(k-1)(d+k-3)} [ 
e_\mu{}^{(1} \xi^{k-1)} - \frac{2}{d+2k-4} g^{(2} \xi^{k-2)}{}_\mu ]
\nonumber \\
\delta_1 \Omega_\mu{}^{a,(k)} &=& \frac{k \alpha_k}{(k-1)(d+k-2)}
 [ e_\mu{}^{(1} \eta^{a,k-1)} - \frac{1}{d+k-3} g^{a(1}
\eta_\mu{}^{k-1)} -  \\
 && \qquad - \frac{2}{d+2k-4} g^{(12} \eta^{a,k-2)}{}_\mu -
\frac{2}{(d+k-3)(d+2k-4)} g^{(12} \eta_\mu{}^{k-2)a} ] \nonumber \\
\delta_1 \Phi_\mu{}^{(k-2)} &=& \alpha_{k-1} \xi_\mu{}^{(k-2)}, \qquad
\delta \Omega_\mu{}^{a,(k-2)} = \frac{k-1}{k-2} \alpha_{k-1} [ 
\eta^{a,(k-2)}{}_\mu + \frac{1}{k-1} \eta_\mu{}^{a(k-2)} ] \nonumber
\end{eqnarray}
provided:
$$
a_{k+1} = b_{k+1} =  \frac{2k}{k-1} \alpha_k
$$
Now, collecting together all corrections for the field $\Phi_k$, we
obtain:
\begin{eqnarray}
\delta_1 \Phi_\mu{}^{(k-1)} &=& \alpha_k \xi_\mu{}^{(k-1)} + 
\frac{(k-1)\alpha_{k-1}}{(k-2)(d+k-4)}  [ e_\mu{}^{(1} \xi^{k-2)} - 
\frac{2}{d+2k-6} g^{(12} \xi^{k-3)}{}_\mu ] \nonumber \\
\delta_1 \Omega_\mu{}^{a,(k-1)} &=& \frac{\alpha_k}{k-1}  [ k
\eta^{a,(k-1)}{}_\mu + \eta_\mu{}^{a(k-1)} ] + \\
 && + \frac{(k-1)\alpha_{k-1}}{(k-2)(d+k-3)}  [ e_\mu{}^{(1}
\eta^{a,k-2)} - \frac{1}{d+k-4} g^{a(1} \eta_\mu{}^{k-2)} - \nonumber
\\
 && - \frac{2}{d+2k-6} g^{(12} \eta^{a,k-3)}{}_\mu +
 \frac{2}{(d+k-4)(d+2k-6)} g^{(12} \eta_\mu{}^{k-3)a} ] \nonumber
\end{eqnarray}
This formulas work for the $k = s$ if we assume $\alpha_s = 0$, while
$k = 1,2$ cases have to be considered separately. Let us collect all
terms in ${\cal L}_1$ containing $\Phi_2$:
$$
 \left\{ \phantom{|}^{\mu\nu}_{ab} \right\} [ - a_3
\Omega_\mu{}^{a,bc} h_\nu{}^c - b_3 \Phi_\mu{}^{ac} \omega_\nu{}^{bc}
+ a_2 \omega_\mu{}^{ab} A_\nu + b_2 h_\mu{}^a F_\nu{}^b ]
$$
Their non-invariance could be compensated by the following corrections
to gauge transformations:
\begin{eqnarray}
\delta_1 \Phi_\mu{}^{ab} &=& \frac{2\alpha_2}{d-1} [ e_\mu{}^{(a}
\xi^{b)} - \frac{2}{d} g^{ab} \xi_\mu ] \nonumber \\
\delta_1 \Omega_\mu{}^{a,bc} &=& \frac{2\alpha_2}{d} [ 
\eta^{a(b} e_\mu{}^{c)}  + \frac{1}{d-1} ( 2 g^{bc}
\eta_\mu{}^a - g^{a(b} \eta_\mu{}^{c)} ) ] \\
\delta_1 A_\mu &=& \alpha_1 \xi_\mu, \qquad
\delta F^{ab} = - 2 \alpha_1 \eta^{ab} \nonumber
\end{eqnarray}
provided:
$$
a_3 = b_3 = 4 \alpha_2, \qquad a_2 = b_2 = \alpha_1
$$
At last, non-invariance of the terms containing $\Phi_1$:
$$
\left\{ \phantom{|}^{\mu\nu}_{ab} \right\} [
 a_2 \omega_\mu{}^{ab} A_\nu + b_2 h_\mu{}^a F_\nu{}^b ]
 - b_1 (A \pi)
$$
could be compensated with:
\begin{equation}
\delta h_\mu{}^a = \frac{\alpha_1}{2(d-2)} e_\mu{}^a \xi, \qquad
\delta \varphi = \alpha_0 \xi, \qquad b_1 = \alpha_0
\end{equation}
We proceed by adding mass-like terms to the Lagrangian:
\begin{equation}
{\cal L}_2 = \sum_{k=2}^s (-1)^k \left\{ \phantom{|}^{\mu\nu}_{ab}
\right\} c_k \Phi_\mu{}^{a(k-2)} \Phi_\nu{}^{b(k-2)} + c_1 h \varphi +
c_0 \varphi^2
\end{equation}
as well as appropriate corrections to gauge transformations:
$$
\delta_2 \Omega_\mu{}^{a,(k)} = \beta_k \left[ k e_\mu{}^a \xi^{(k)}
- e_\mu{}^{(1} \xi^{k-1)a} - \frac{1}{d+k-3} [ (k-1) g^{a(1}
\xi_\mu{}^{k-1)} - 2 g^{(12} \xi_\mu{}^{k-2)a} ] \right], \quad k > 1
$$
\begin{equation}
\delta_2 \omega_\mu{}^{ab} = \beta_1 [e_\mu{}^a \xi^b - e_\mu{}^b
\xi^a], \qquad \delta_2 \pi^a = \beta_0 \xi^a
\end{equation}
Cancellation of variations at order $m^2$ (taking into account
non-invariance of kinetic terms due to non-commutativity of covariant
derivatives) requires:
$$
c_k = - \frac{k^2(d+2k-2)(d+k-4)}{(k-1)(d+k-3)(d+2k-4)} 
\alpha_k{}^2 + \frac{k(k-1)}{k-2} \alpha_{k-1}{}^2  - k(d+k-4)
\kappa, \quad k > 2
$$
$$
c_2 = - \frac{4(d^2-4)}{d(d-1)} \alpha_2{}^2 + \alpha_1{}^2 - 2(d-2)
\kappa \qquad c_1 = - \alpha_1 \alpha_0
$$
$$
\beta_k = - \frac{c_{k+1}}{(k+1)(d-2)}, \quad k > 0, \qquad 
\beta_0 = c_1
$$
At last, all variations at order $m^3$ cancel provided:
$$
k (d+k-2) c_{k+1} = (k-1) (d+k-3) c_k, \quad k > 1
$$
$$
4 (d-1) c_2 = (d-2) \alpha_0{}^2, \qquad
c_0 = \frac{d}{d-2} \alpha_1{}^2
$$
These last relations give us recurrent relations on the main parameters
$\alpha$:
$$
\frac{(k+1)^2(d+2k)}{d+2k-2} \alpha_{k+1}{}^2
- \frac{2k^3(d+2k-3)}{(k-1)(d+2k-4)} \alpha_k{}^2
+ \frac{k(k-1)^2}{k-2} \alpha_{k-1}{}^2 +
2k(d+2k-3) \kappa = 0
$$
To solve these relations, recall that $\alpha_s = 0$, while our
definition of mass gives in this case $\alpha_{s-1}{}^2 = 
\frac{s-2}{(s-1)^2} m^2$. We obtain:
$$
\alpha_k{}^2 = \frac{(k-1)(s-k)(d+s+k-3)}{k^2(d+2k-2)} [ m^2 -
(s-k-1)(d+s+k-4) \kappa ], \quad k > 1
$$
$$
\alpha_1{}^2 = \frac{2(s-1)(d+s-2)}{d} [ m^2 - (s-2) (d+s-3) \kappa ]
$$
$$
\alpha_0{}^2 = \frac{4s(d+s-3)}{(d-2)} [ m^2 - (s-1)(d+s-4) \kappa ]
$$
Using this solution we get final expressions for parameters $c$:
$$
c_k = \frac{s(d+s-3)}{(k-1)(d+k-3)} [ m^2 - (s-1)(d+s-4) \kappa ],
\quad k > 1
$$
$$
c_1 = - \alpha_1 \alpha_0, \qquad c_0 = \frac{d}{d-2} \alpha_1{}^2
$$
Collecting all results we obtain final Lagrangian for gauge invariant
description of massive spin $s$ particles in $(A)dS_d$ space:
\begin{eqnarray}
{\cal L} &=& \sum_{k=2}^s {\cal L}_0(\Phi_k) + \frac{1}{4} F_{ab}{}^2
- \frac{1}{2} \left\{ \phantom{|}^{\mu\nu}_{ab} \right\} F^{ab} D_\mu
A_\nu - \frac{1}{2} \pi_a{}^2 + \left\{ \phantom{|}^{\mu}_{a} \right\}
\pi^a D_\mu \varphi + \nonumber \\
 && + \sum_{k=2}^s (-1)^k \left\{ \phantom{|}^{\mu\nu}_{ab}
\right\} a_k [ \Omega_\mu{}^{a,b(k-2)} \Phi_\nu{}^{(k-2)} + 
\Phi_\mu{}^{a(k-2)} \Omega_\nu{}^{b,(k-2)} ] - \alpha_0 (A \pi) +
\nonumber \\
 && + \sum_{k=2}^s (-1)^k \left\{ \phantom{|}^{\mu\nu}_{ab}
\right\} c_k \Phi_\mu{}^{a(k-2)} \Phi_\nu{}^{b(k-2)} - \alpha_1
\alpha_0 h \varphi + \frac{d}{d-2} \alpha_1{}^2 \varphi^2
\end{eqnarray}
$$
a_k = \frac{2(k-1)}{k-2} \alpha_{k-1}, \quad k > 2, \quad a_2 =
\alpha_1, \quad c_k =  \frac{s(d+s-3)}{(k-1)(d+k-3)} M^2
$$
where $M^2 = m^2 - (s-1)(d+s-4) \kappa$. As for the gauge
transformations leaving this Lagrangian invariant, we have seen that
the most complicated part is related with gauge transformations of
auxiliary fields $\Omega$. So we reproduce here gauge transformations
for physical fields only:
\begin{eqnarray}
\delta \Phi_\mu{}^{(k)} &=& D_\mu \xi^{(k)} + \eta_\mu{}^{(k)} +
\alpha_{k+1} \xi_\mu{}^{(k)} +
\frac{k \alpha_k}{(k-1)(d+k-3)} [ e_\mu{}^{(1} \xi^{k-1)} -
\frac{2}{d+2k-4} g^{(2} \xi^{k-2)}{}_\mu ]  \nonumber \\
\delta h_\mu{}^a &=& D_\mu \xi^a + \eta_\mu{}^a + \alpha_2 \xi_\mu{}^a
+ \frac{\alpha_1}{2(d-2)} e_\mu{}^a \xi, \quad
\delta A_\mu = D_\mu \xi + \alpha_1 \xi_\mu, \quad \delta \varphi =
\alpha_0 \xi
\end{eqnarray}
As in the spin 2 and spin 3 cases, we see that massless limit is
possible in $AdS$ (and Minkowski) space only. In this, massive spin
$s$ particle decompose into massless spin $s$ and massive spin $s-1$
ones. In $dS$ space we again find unitary forbidden region $m^2 <
(s-1)(d+s-4) \kappa$. At the boundary of this region  $m^2 =
(s-1)(d+s-4) \kappa$ scalar field decouples and we obtain first
partially massless theory (note that in this case all explicit
mass-like terms vanish). Inside this forbidden region we obtain a
number of partially massless theories. Namely, if one of the $\alpha_k
= 0$, then all fields $\Phi_l$ with $0 \le l \le k$ decouple, while
the remaining fields with $k+1 \le l \le s$ give gauge invariant
description of corresponding partially massless particle. Let us give
here only one concrete example --- the most simple one where only two
fields $\Phi_s$ and $\Phi_{s-1}$ remain. It happens when $m^2 =
(d+2s-6) \kappa$. Corresponding Lagrangian looks as:
\begin{eqnarray}
{\cal L} &=& {\cal L}_0 (\Phi_s) + {\cal L}_0 (\Phi_{s-1}) +  (-1)^s
\frac{2m}{\sqrt{s-2}} \left\{ \phantom{|}^{\mu\nu}_{ab} \right\}  [
\Omega_\mu{}^{a,b(s-2)} \Phi_\nu{}^{(s-2)} +  \Phi_\mu{}^{a(s-2)}
\Omega_\nu{}^{b,(s-2)} ]  - \nonumber \\
 && - (-1)^s s (d+s-5) \kappa \left\{ \phantom{|}^{\mu\nu}_{ab}
\right\} [ \frac{s-2}{s-1} \Phi_\mu{}^{a(s-2)} \Phi_\nu{}^{b(s-2)} - 
\frac{d+s-3}{d+s-4}  \Phi_\mu{}^{a(s-3)} \Phi_\nu{}^{b(s-3)} ]
\end{eqnarray}
while gauge transformations leaving it invariant have the form:
\begin{eqnarray}
\delta \Phi_\mu{}^{(s-1)} &=& D_\mu \xi^{(s-1)} + \eta_\mu{}^{(s-1)} +
\frac{m}{\sqrt{s-2}(d+s-4)} [ e_\mu{}^{(1} \xi^{s-2)} -
\frac{2}{d+2s-6} g^{(2} \xi^{s-3)}{}_\mu ]  \nonumber \\
\delta \Phi_\mu{}^{(s-2)} &=& D_\mu \xi^{(s-2)} + \eta_\mu{}^{(s-2)} +
\frac{m\sqrt{s-2}}{s-1} \xi_\mu{}^{(s-2)} 
\end{eqnarray}

\section{Spin 5/2}

In this and in the next section we will work in four-dimensional
space-time assuming that all fermionic objects are Majorana ones,
however all results could be easily generalized to arbitrary dimension
with appropriate spinors. We start here with the first non-trivial
example --- spin 5/2. For the description of free massless spin 5/2
particles \cite{Vas88} there is no need to introduce any auxiliary
fields, though they play important role for construction of
interactions. So the only object we need --- spin-tensor
$\Psi_\mu{}^a$ which is $\gamma$-transverse $\gamma^a \Psi_\mu{}^a =
0$. To describe correct number of physical degrees of freedom, theory
must be invariant under the following gauge transformations:
\begin{equation}
\delta \Psi_\mu{}^a = \partial_\mu \xi^a + \eta_\mu{}^a, \quad
\gamma^a \xi^a = 0, \quad \eta^{ab} = - \eta^{ba}, \quad
\gamma^a \eta^{ab} = 0
\end{equation}
It is not hard to construct gauge invariant Lagrangian describing
massless particle:
\begin{equation}
{\cal L}_0 = \frac{i}{2} \left\{ \phantom{|}^{\mu\nu\alpha}_{abc}
\right\} [ \bar{\Psi}_\mu{}^d \gamma^a \gamma^b \gamma^c \partial_\nu
\Psi_\alpha{}^d - 6 \bar{\Psi}_\mu{}^a \gamma^b \partial_\nu
\Psi_\alpha{}^c ]
\end{equation}
where relative coefficient is determined by the $\eta$-invariance. Let
us first of all consider deformation to $AdS$ space. Working with
covariant derivatives one has to take into account (implicit) spinor
indices on fermionic objects, e.g.:
\begin{equation}
 [ D_\mu, D_\nu ] \xi^a = - \kappa ( e_\mu{}^a \xi_\nu - e_\nu{}^a
\xi_\mu + \frac{1}{2} \sigma_{\mu\nu} \xi^a )
\end{equation}
As is well known, in $AdS$ space gauge transformations has to be
modified:
\begin{equation}
\delta \Psi_\mu{}^a = D_\mu \xi^a + i \alpha_0 \gamma_\mu \xi^a +
\eta_\mu{}^a, \qquad 
\end{equation}
and, for this transformations to be compatible with the constraint 
$\gamma^a \Psi_\mu{}^a = 0$, the constraint on $\eta$ parameter also
has to be changed:
$$
\gamma^a \eta^{ab} = 0 \quad \Rightarrow \quad
\gamma^a \eta^{ab} = 2 i \alpha_0 \xi^b
$$
Non-invariance of (covariantized) massless Lagrangian under new gauge
transformations (taking into account contribution from
$\eta$-transformations due to constraint) has the form:
$$
\delta_\xi {\cal L}_0 = - 12 \alpha_0  \left\{
\phantom{|}^{\mu\nu}_{ab} \right\} [ \bar{\Psi}_\mu{}^c \gamma^a
\gamma^b D_\nu \xi^c + 2 \bar{\Psi}_\mu{}^a D_\nu \xi^b ] + 30 i
\kappa (\bar{\Psi} \gamma)^a \xi^a
$$
where $ (\bar{\Psi} \gamma)^a = \bar{\Psi}_\mu{}^a \gamma^\mu$ and
could be compensated by adding to the Lagrangian mass-like terms:
$$
{\cal L}_1 = \left\{ \phantom{|}^{\mu\nu}_{ab} \right\} [ a_1
\Psi_\mu{}^c \gamma^a \gamma^b \Psi_\nu{}^c + a_2 \Psi_\mu{}^a
\Psi_\nu{}^b ]
$$
provided $a_1 = 6 \alpha_0$, $a_2 = 12 \alpha_0$,  
$\alpha_0{}^2 = - \frac{\kappa}{4}$. Thus, the Lagrangian for massless
spin 5/2 particle and gauge transformations leaving it invariant have
the form:
\begin{eqnarray*}
{\cal L} &=& \frac{i}{2} \left\{ \phantom{|}^{\mu\nu\alpha}_{abc}
\right\} [ \bar{\Psi}_\mu{}^d \gamma^a \gamma^b \gamma^c D_\nu
\Psi_\alpha{}^d - 6 \bar{\Psi}_\mu{}^a \gamma^b D_\nu
\Psi_\alpha{}^c ] + \\
 && + 3 \sqrt{-\kappa} \left\{ \phantom{|}^{\mu\nu}_{ab} \right\} [
\Psi_\mu{}^c \gamma^a \gamma^b \Psi_\nu{}^c + 2 \Psi_\mu{}^a
\Psi_\nu{}^b ]
\end{eqnarray*}
\begin{equation}
\delta \Psi_\mu{}^a = D_\mu \xi^a + i \frac{\sqrt{-\kappa}}{2}
\gamma_\mu \xi^a + \eta_\mu{}^a, \qquad 
\gamma^a \eta^{ab} =  i \sqrt{-\kappa} \xi^b
\end{equation}
Note that working with massless fermions in $AdS$ space very often one
introduces a convenient generalized covariant derivative, e.g.
$$
\nabla_\mu \xi^a = D_\mu \xi^a + i \frac{\sqrt{-\kappa}}{2}
\gamma_\mu \xi^a
$$
As we will see later on, for massive case parameter $\alpha_0$ will
depends both on mass and cosmological constant, so we will not
introduce such covariant derivative here.

Let us turn to the massive case. Gauge invariant description of
massive spin 5/2 particle \cite{Met06} requires introduction of two
additional fields with spin 3/2 and spin 1/2. We consider a Lagrangian
which is a sum of covariantized kinetic terms for all three fields
plus all possible mass-like terms:
\begin{eqnarray}
{\cal L} &=& \frac{i}{2} \left\{ \phantom{|}^{\mu\nu\alpha}_{abc}
\right\} [ \bar{\Psi}_\mu{}^d \gamma^a \gamma^b \gamma^c D_\nu
\Psi_\alpha{}^d - 6 \bar{\Psi}_\mu{}^a \gamma^b D_\nu
\Psi_\alpha{}^c  - \bar{\psi}_\mu \gamma^a \gamma^b \gamma^c D_\nu
\psi_\alpha ] + \frac{i}{2} \bar{\chi} \hat{D} \chi + \nonumber \\
 && + \left\{ \phantom{|}^{\mu\nu}_{ab} \right\} [ a_1
\Psi_\mu{}^c \gamma^a \gamma^b \Psi_\nu{}^c + 2 a_1 \Psi_\mu{}^a
\Psi_\nu{}^b + i a_2 \bar{\Psi}_\mu{}^a \gamma^b \psi_\nu +
a_3 \bar{\psi}_\mu \gamma^a \gamma^b \psi_\nu ] + \nonumber \\
 && + i a_4 (\bar{\psi}\gamma) \chi + a_5 \bar{\chi} \chi
\end{eqnarray}
as well as the most general form of corresponding gauge
transformations:
\begin{eqnarray}
\delta \Psi_\mu{}^a &=& D_\mu \xi^a + i \alpha_1 \gamma_\mu \xi^a +
\alpha_2 e_\mu{}^a \xi + \eta_\mu{}^a,  \qquad 
(\gamma \eta)^a = 2 i \alpha_1 \xi^a + \alpha_2 \gamma^a \xi
\nonumber \\
\delta \psi_\mu &=& D_\mu \xi + i \alpha_3 \gamma_\mu \xi + \alpha_4
\xi_\mu, \qquad \delta \chi = \alpha_5 \xi
\end{eqnarray}
Note that the constraint on $\eta$ parameter was changed.  First of
all, the requirement that Lagrangian has to be invariant under such
gauge transformations allows one to express all parameters in the
Lagrangian and gauge transformations in terms of three main parameters
$\alpha_1$, $\alpha_4$ and $\alpha_5$:
$$
a_1 = 6 \alpha_1, \quad a_2 = 6 \alpha_4, \quad a_3 = - 9 \alpha_1,
\quad a_4 = \alpha_5, \quad a_5 = - 6 \alpha_1, \quad
\alpha_2 = \frac{\alpha_4}{4}, \quad \alpha_3 = 3 \alpha_1
$$
Already from these formulas we see that it is the limit $\alpha_4
\rightarrow 0$ that corresponds to the massless limit. Our usual
convention on the mass normalization gives here $\alpha_4{}^2 = 5
m^2/4$. In this, two other main parameters are determined by the
relations:
\begin{equation}
16 \alpha_1{}^2 = m^2 - 4 \kappa, \qquad \qquad 
\alpha_5{}^2 = 24 (m^2 - 3 \kappa)
\end{equation}
So the resulting Lagrangian and gauge transformations could be written
as follows:
\begin{eqnarray}
{\cal L} &=& \frac{i}{2} \left\{ \phantom{|}^{\mu\nu\alpha}_{abc}
\right\} [ \bar{\Psi}_\mu{}^d \gamma^a \gamma^b \gamma^c D_\nu
\Psi_\alpha{}^d - 6 \bar{\Psi}_\mu{}^a \gamma^b D_\nu
\Psi_\alpha{}^c  - \bar{\psi}_\mu \gamma^a \gamma^b \gamma^c D_\nu
\psi_\alpha ] + \frac{i}{2} \bar{\chi} \hat{D} \chi + \nonumber \\
 && + \left\{ \phantom{|}^{\mu\nu}_{ab} \right\} [ 6 \alpha_1
\Psi_\mu{}^c \gamma^a \gamma^b \Psi_\nu{}^c + 12 \alpha_1 \Psi_\mu{}^a
\Psi_\nu{}^b + i \frac{3\sqrt{5}}{2} m \bar{\Psi}_\mu{}^a \gamma^b
\psi_\nu - 9 \alpha_1 \bar{\psi}_\mu \gamma^a \gamma^b \psi_\nu ] +
\nonumber \\
 && + i \alpha_5 (\bar{\psi}\gamma) \chi - 6 \alpha_1 \bar{\chi} \chi
\end{eqnarray}
\begin{eqnarray}
\delta \Psi_\mu{}^a &=& D_\mu \xi^a + i \alpha_1 \gamma_\mu \xi^a +
\frac{\sqrt{5}}{8} m e_\mu{}^a \xi + \eta_\mu{}^a \nonumber \\
\delta \psi_\mu &=& D_\mu \xi + 3 i \alpha_1 \gamma_\mu \xi +
\frac{\sqrt{5}}{2} m \xi_\mu, \qquad \delta \chi = \alpha_5 \xi
\end{eqnarray}
From these results we see that massless limit is once again possible
in $AdS$ space ($\kappa < 0$) only, in this the whole system
decomposes into massless spin 5/2 particle and massive spin 3/2 one.
In $dS$ space we also obtain unitary forbidden region $m^2 < 4
\kappa$. At the boundary of this region, i.e. $\alpha_1 = 0$, the
theory greatly simplifies:
\begin{eqnarray}
{\cal L} &=& \frac{i}{2} \left\{ \phantom{|}^{\mu\nu\alpha}_{abc}
\right\} [ \bar{\Psi}_\mu{}^d \gamma^a \gamma^b \gamma^c D_\nu
\Psi_\alpha{}^d - 6 \bar{\Psi}_\mu{}^a \gamma^b D_\nu
\Psi_\alpha{}^c  - \bar{\psi}_\mu \gamma^a \gamma^b \gamma^c D_\nu
\psi_\alpha ] + \frac{i}{2} \bar{\chi} \hat{D} \chi + \nonumber \\
 && +  i \frac{3\sqrt{5}}{2} m \left\{ \phantom{|}^{\mu\nu}_{ab}
\right\} \bar{\Psi}_\mu{}^a \gamma^b \psi_\nu 
 + i \sqrt{6} m (\bar{\psi}\gamma) \chi
\end{eqnarray}
\begin{equation}
\delta \Psi_\mu{}^a = D_\mu \xi^a  + \frac{\sqrt{5}}{8} m e_\mu{}^a
\xi + \eta_\mu{}^a, \qquad
\delta \psi_\mu = D_\mu \xi  + \frac{\sqrt{5}}{2} m \xi_\mu, \qquad
\delta \chi = \sqrt{6} m \xi
\end{equation}
Inside the forbidden region there is a special value $m^2 = 3 \kappa$
when $\alpha_5 = 0$. In this case spinor field decouples, while two
other give gauge invariant description of partially massless particle
(with helicities $\pm 5/2, \pm 3/2$) with the Lagrangian:
\begin{eqnarray}
{\cal L} &=& \frac{i}{2} \left\{ \phantom{|}^{\mu\nu\alpha}_{abc}
\right\} [ \bar{\Psi}_\mu{}^d \gamma^a \gamma^b \gamma^c D_\nu
\Psi_\alpha{}^d - 6 \bar{\Psi}_\mu{}^a \gamma^b D_\nu
\Psi_\alpha{}^c  - \bar{\psi}_\mu \gamma^a \gamma^b \gamma^c D_\nu
\psi_\alpha ] + \nonumber \\
 && + \left\{ \phantom{|}^{\mu\nu}_{ab} \right\} [ 6 \alpha_1
\Psi_\mu{}^c \gamma^a \gamma^b \Psi_\nu{}^c + 12 \alpha_1 \Psi_\mu{}^a
\Psi_\nu{}^b + i \frac{3\sqrt{5}}{2} m \bar{\Psi}_\mu{}^a \gamma^b
\psi_\nu - 9 \alpha_1 \bar{\psi}_\mu \gamma^a \gamma^b \psi_\nu ] 
\end{eqnarray}
which is invariant under the following gauge transformations:
\begin{eqnarray}
\delta \Psi_\mu{}^a &=& D_\mu \xi^a + i \alpha_1 \gamma_\mu \xi^a +
\frac{\sqrt{5}}{8} m e_\mu{}^a \xi + \eta_\mu{}^a \nonumber \\
\delta \psi_\mu &=& D_\mu \xi + 3 i \alpha_1 \gamma_\mu \xi +
\frac{\sqrt{5}}{2} m \xi_\mu
\end{eqnarray}
However, in this case $\alpha_1{}^2 = - \kappa/16 < 0$.

\section{Arbitrary half-integer spin}

For the description of massless spin $s+\frac{1}{2}$ particle ($s
=1,2 \dots$) one needs \cite{Vas88} spin-tensor $\Psi_\mu{}^{a_1 \dots
a_{s-1}}$ symmetric on local indices and satisfying a constraint
$\gamma^{a_1} \Psi_\mu{}^{a_1 \dots a_{s-1}} = 0$. In this section we
will use the same condensed notations as before, e.g. our main field
will be denoted as $\Psi_\mu{}^{(s-1)}$. Free massless Lagrangian has
to be invariant under the following gauge transformations:
\begin{equation}
\delta \Psi_\mu{}^{(s-1)} = \partial_\mu \xi^{(s-1)} +
\eta_\mu{}^{(s-1)}, \quad \gamma^a \xi^{a(s-2)} = 0, \quad \gamma^a
\eta^{a,(s-1)} = \gamma^b \eta^{a,b(s-2)} = 0, \quad 
\eta^{(a,s-1)} = 0
\end{equation}
Such a Lagrangian could be written in the following form:
\begin{equation}
(-1)^{s} {\cal L}_0 = \frac{i}{2} \left\{
\phantom{|}^{\mu\nu\alpha}_{abc} \right\} [ \bar{\Psi}_\mu{}^{(s-1)}
\gamma^a \gamma^b \gamma^c \partial_\nu \Psi_\alpha{}^{(s-1)} - 6(s-1)
\bar{\Psi}_\mu{}^{a(s-2)} \gamma^b \partial_\nu \Psi_\alpha{}^{c(s-2)}
]
\end{equation}
In order to describe massless particle in $AdS$ space one has first of
all change gauge transformations and constraint on $\eta$ parameter:
\begin{equation}
\delta \Psi_\mu{}^{(s-1)} = D_\mu \xi^{(s-1)} + i \alpha_0
\gamma_\mu \xi^{(s-1)} + \eta_\mu{}^{(s-1)},
\qquad (\gamma \eta)^{(s-1)} = 2i(s-1) \alpha_0 \xi^{(s-1)}
\end{equation}
and supplement the Lagrangian with additional mass-like terms of the
form:
\begin{equation}
(-1)^{s} {\cal L}_1 = 3s \alpha_0 \left\{ \phantom{|}^{\mu\nu}_{ab}
\right\} [ \bar{\Psi}_\mu{}^{(s-1)} \gamma^a \gamma^b
\Psi_\nu{}^{(s-1)} + 2 (s-1) \bar{\Psi}_\mu{}^{a(s-2)}
\Psi_\nu{}^{b(s-2)} ]
\end{equation}
Resulting Lagrangian will be gauge invariant provided $\alpha_0{}^2 =
- \frac{\kappa}{4}$.

Now let us turn to the massive case. To construct gauge invariant
description of massive spin $s+\frac{1}{2}$ particle one needs
\cite{Met06} a set of fields with spins $s+\frac{1}{2}$, 
$s-\frac{1}{2}$, ... $\frac{1}{2}$. Thus we introduce fields
$\Psi_\mu{}^{(k)}$, $0 \le k \le s-1$ and spinor $\chi$. Consider a
Lagrangian which is a sum of covariantized kinetic terms for all
fields plus the most general mass-like terms:
\begin{eqnarray}
{\cal L} &=& \sum_{k=0}^{s-1} (-1)^{k+1} \frac{i}{2}
\left\{ \phantom{|}^{\mu\nu\alpha}_{abc} \right\} [
\bar{\Psi}_\mu{}^{(k)} \gamma^a \gamma^b \gamma^c D_\nu
\Psi_\alpha{}^{(k)} - 6k \bar{\Psi}_\mu{}^{a(k-1)} \gamma^b D_\nu
\Psi_\alpha{}^{c(k-1)} ] + \frac{i}{2} \bar{\chi} \hat{D} \chi  +
\nonumber \\
 && \sum_{k=1}^{s-1} (-1)^{k+1} 
\left\{ \phantom{|}^{\mu\nu}_{ab} \right\} [ a_k
\bar{\Psi}_\mu{}^{(k)} \gamma^a \gamma^b \Psi_\nu{}^{(k)} + 2ka_k
\bar{\Psi}_\mu{}^{a(k-1)} \Psi_\nu{}^{b(k-1)} + i b_k
\bar{\Psi}_\mu{}^{a(k-1)} \gamma^b \Psi_\nu{}^{(k-1)} ] - \nonumber \\
 && \qquad \quad - a_0 \left\{ \phantom{|}^{\mu\nu}_{ab} \right\}
\bar{\Psi}_\mu \gamma^a \gamma^b \Psi_\nu  + i b_0 (\bar{\Psi} \gamma)
\chi  + c_0 \bar{\chi} \chi
\end{eqnarray}
as well as the following ansatz for the gauge transformations (note
the modification of constraints on $\eta$ parameters):
$$
\delta \Psi_\mu{}^{(k)} = D_\mu \xi^{(k)} + \alpha_{k+1}
\xi_\mu{}^{(k)} + i \beta_k \gamma_\mu \xi^{(k)} + \rho_k [
e_\mu{}^{(1} \xi^{k-1)} - \frac{1}{k} g^{(12} \xi^{k-2)}{}_\mu ] +
\eta_\mu{}^{(k)}
$$ 
\begin{equation}
\delta \chi = \alpha_0 \xi, \qquad (\gamma \eta)^{(k)} = 2ik\beta_k
\xi^{(k)} + \frac{1}{k} \rho_k \gamma^{(1} \xi^{k-1)}
\end{equation}
Variations of order $m$ give:
$$
a_k = 3(k+1) \beta_k, \qquad b_k = 6k \alpha_k, \qquad
\rho_k = \frac{k \alpha_k}{(k+1)^2}, \qquad b_0 = \alpha_0
$$
while variations of order $m^2$ (including contributions form the
commutators of covariant derivatives) cancel provided:
$$
(k+1) \beta_k = (k+3) \beta_{k+1}
$$
$$
4(k+1)(2k+3) \beta_k{}^2 - 2k \alpha_k{}^2 + \frac{2(k+1)^2}{k+2}
\alpha_{k+1}{}^2 + (2k+3)(k+1) \kappa = 0
$$
$$
\alpha_0{}^2 = 3 \alpha_1{}^2 + 36 \beta_0{}^2 + 9 \kappa, \qquad c_0
= - 2 \beta_0
$$
To solve these relations we proceed as follows. First of all, noting
that $\alpha_s = 0$ while our definition of mass gives this time
$\alpha_{s-1}{}^2 = \frac{(2s+1)m^2}{2s(s-1)}$, from the second
relation with $k = s-1$ we obtain:
$$
\beta_{s-1}{}^2 = \frac{m^2 - s^2 \kappa}{4s^2}
$$
Then solving the first relation recurrently we get all other $\beta$-s:
$$
\beta_k{}^2 = \frac{(s+1)^2 (m^2 - s^2 \kappa)}{4(k+1)^2(k+2)^2}
$$
Now the second relation becomes recurrent relation on $\alpha$-s and
can easily be solved. Result:
$$
\alpha_k{}^2 = \frac{(s-k)(s+k+2)}{2k(k+1)} [ m^2 - (s^2 - (k+1)^2)
\kappa ]
$$
$$
\alpha_0{}^2 = 3s(s+2) [ m^2 - (s^2 - 1) \kappa ]
$$
The final Lagrangian and gauge transformations in terms of $\alpha$
and $\beta$ look as follows:
\begin{eqnarray}
{\cal L} &=& \sum_{k=0}^{s-1} (-1)^{k+1} \frac{i}{2}
\left\{ \phantom{|}^{\mu\nu\alpha}_{abc} \right\} [
\bar{\Psi}_\mu{}^{(k)} \gamma^a \gamma^b \gamma^c D_\nu
\Psi_\alpha{}^{(k)} - 6k \bar{\Psi}_\mu{}^{a(k-1)} \gamma^b D_\nu
\Psi_\alpha{}^{c(k-1)} ] + \frac{i}{2} \bar{\chi} \hat{D} \chi  +
\nonumber \\
 && \sum_{k=1}^{s-1} (-1)^{k+1} 
\left\{ \phantom{|}^{\mu\nu}_{ab} \right\} [ 3(k+1) \beta_k
\bar{\Psi}_\mu{}^{(k)} \gamma^a \gamma^b \Psi_\nu{}^{(k)} + 6k (k+1)
\beta_k \bar{\Psi}_\mu{}^{a(k-1)} \Psi_\nu{}^{b(k-1)} + \nonumber \\
 && \qquad +  6i k \alpha_k \bar{\Psi}_\mu{}^{a(k-1)}
\gamma^b \Psi_\nu{}^{(k-1)} ] - 3 \beta_0
\left\{ \phantom{|}^{\mu\nu}_{ab} \right\}
\bar{\Psi}_\mu \gamma^a \gamma^b \Psi_\nu
 + i \alpha_0 (\bar{\Psi} \gamma) \chi  -
2 \beta_0 \bar{\chi} \chi
\end{eqnarray}
$$
\delta \Psi_\mu{}^{(k)} = D_\mu \xi^{(k)} + \alpha_{k+1}
\xi_\mu{}^{(k)} + i \beta_k \gamma_\mu \xi^{(k)} + 
\frac{k\alpha_k}{(k+1)^2} [ e_\mu{}^{(1} \xi^{k-1)} - \frac{1}{k}
g^{(12}
\xi^{k-2)}{}_\mu ] + \eta_\mu{}^{(k)}
$$
\begin{equation}
\delta \chi = \alpha_0 \xi, \qquad (\gamma \eta)^{(k)} = 2ik\beta_k
\xi^{(k)} + \frac{2 \alpha_k}{k(k+1)^2} \gamma^{(1} \xi^{k-1)}
\end{equation}

Now we are ready to analyze main properties of the theory obtained. It
is hardly come as a surprise that massless limit turns out to be
possible in $AdS$ space only, in this massive spin $s+\frac{1}{2}$
particle decompose into massless spin $s+\frac{1}{2}$ and massive
$s-\frac{1}{2}$ ones. In $dS$ space we once again find unitary
forbidden region $m^2 < s^2 \kappa$. At the boundary of this region
all parameters $\beta$ become zero and theory greatly simplifies:
\begin{eqnarray}
{\cal L} &=& \sum_{k=0}^{s-1} (-1)^{k+1} \frac{i}{2}
\left\{ \phantom{|}^{\mu\nu\alpha}_{abc} \right\} [
\bar{\Psi}_\mu{}^{(k)} \gamma^a \gamma^b \gamma^c D_\nu
\Psi_\alpha{}^{(k)} - 6k \bar{\Psi}_\mu{}^{a(k-1)} \gamma^b D_\nu
\Psi_\alpha{}^{c(k-1)} ] + \frac{i}{2} \bar{\chi} \hat{D} \chi  +
\nonumber \\
 && \sum_{k=1}^{s-1} (-1)^{k+1} 6i k \alpha_k
\left\{ \phantom{|}^{\mu\nu}_{ab} \right\}
\bar{\Psi}_\mu{}^{a(k-1)} \gamma^b \Psi_\nu{}^{(k-1)} 
 + i \alpha_0 (\bar{\Psi} \gamma) \chi 
\end{eqnarray}
$$
\delta \Psi_\mu{}^{(k)} = D_\mu \xi^{(k)} + \alpha_{k+1}
\xi_\mu{}^{(k)}  + \frac{k\alpha_k}{(k+1)^2} [ e_\mu{}^{(1} \xi^{k-1)}
- \frac{1}{k} g^{(12} \xi^{k-2)}{}_\mu ] + \eta_\mu{}^{(k)}
$$
\begin{equation}
\delta \chi = \alpha_0 \xi, \qquad (\gamma \eta)^{(k)} = 
\frac{2 \alpha_k}{k(k+1)^2} \gamma^{(1} \xi^{k-1)}
\end{equation}
Note the essential difference between integer and half-integer
cases \cite{Gar03}: for the integer spin at the boundary of unitary
forbidden region spin 0 field decouples and we obtain first partially
massless theory, while for the half-integer spin all the partially
massless theories "live" inside the forbidden region. Indeed, for any
value of mass where one of the parameters $\alpha_k$ becomes zero, all
fields with spins $l+\frac{1}{2}$ for $0 \le l \le k$ decouple, while
remaining fields describe partially massless theory.

\end{document}